**Hydrothermal co-liquefaction of synthetic polymers and *Miscanthus Giganteus;* synergistic and antagonistic effects**

Juliano Souza dos Passos[1,3], Marianne Glasius[2,3], Patrick Biller[1,3]*
1 - Biological and Chemical Engineering, Aarhus University, Hangøvej 2, DK-8200 Aarhus N, Denmark
2 - Department of Chemistry, Aarhus University, Langelandsgade 140, DK-8000 Aarhus C, Denmark
3 - Aarhus University Centre for Circular Bioeconomy, Blichers Allé 20, DK-8830, Tjele, Denmark
*Corresponding author: pbiller@eng.au.dk

**ABSTRACT**

Synthetic polymers constitute one of the main carbon-containing wastes generated nowadays. In this study, combined hydrothermal liquefaction (co-HTL) is evaluated for 1:1 mixtures of *Miscanthus Giganteus* and different synthetic polymers – including poly-acrylonitrile-butadiene-styrene (ABS), Bisphenol-A based Epoxy resin, high density polyethylene (HDPE), low density polyethylene (LDPE), polyamide 6 (PA6), polyamide 6/6 (PA66), polyethylene terephthalate (PET), polycarbonate (PC), polypropylene (PP), polystyrene (PS) and polyurethane foam (PUR) – using batch HTL reactors at 350 °C. Based on mass balances and oil composition, a comprehensive discussion of observed interactions is presented. The results show that even though polyolefins do not depolymerize under these conditions, the oil products depicts that these materials interact with miscanthus oil changing its composition. Bisphenol-A based polymers as PC and epoxy resins both contribute for the formation of monomer-like structures in the oil. PET increases the presence of carboxyl groups, while polyamides and PUR increase significantly the oil yield, modifying oil composition towards nitrogen-containing molecules. PUR co-HTL was found to increase both oil carbon and energy yields, leading to process improvement when compared to pure miscanthus processing.

## 1. INTRODUCTION

Out of the 8,300 million metric tons of virgin plastics produced up to 2017, a mere 7% was recycled and another 12% burned for energy recovery, leaving a little more than 6,500 million metric tons of waste accumulated in the environment.[1] A large fraction of this waste accumulation takes place in an uncontrolled manner, leading to mixing of different waste streams with synthetic resins. Given the total synthetic polymer production is foreseen to double in the next 20 years accompanied by an increase of gross domestic product in developing countries[2], accumulation of plastic waste is also expected to increase. Technologies that can recycle mixed wastes comprised of both synthetic and biological polymers are hence highly desired in order to reuse the fossil derived carbon contained in plastic wastes, beyond incineration or landfilling.

Hydrothermal liquefaction (HTL) is a technology that uses the properties of near-critical water to depolymerize carbon-containing materials into a crude-like substance often referred to as biocrude.[3] The technology has shown the potential to convert materials such as lignocellulosics, organic wastes[4,5] and algae[6] both individually and mixed[7,8]. Recently, co-liquefaction of wastes has been receiving increasing attention as researchers observe synergies in the liquefaction behavior of different organic wastes and biomass, depicting opportunities to ease processing and increase process efficiency. For instance, Loblolly Pine co-liquefaction with both manure and sewage sludge has been reported to improve biocrude yields and quality.[9] Furthermore, sewage sludge co-liquefaction with a variety of lignocellulosic materials has also shown potential to improve process efficiency.[10]

Synthetic polymers have received less attention in the HTL literature, partially due to the fact that pyrolysis is often the preferred route for chemical recycling of polyolefins and polystyrene (PS). This technology is carried out above the thermal cracking temperature of these synthetic materials (~500°C), which account for the majority of synthetic wate.[1] A recent study by the authors[11] on subcritical HTL of synthetic polymers (350°C) has shown that only some polymer resins are suitable for oil or monomer recovery via HTL. Polyolefins and PS, for instance, were shown not to decompose at these conditions, while polymers with heteroatoms (N or O) in the backbone appeared more promising. Since different biomass fractions have previously shown synergies in HTL and avoided the need for catalysts, it is worth investigating if similar synergies exist between biomass and synthetic polymers. The combination of biomass and synthetic wastes via HTL has the potential to improve the conversion of polymers and the overall carbon throughput and recovery, hence, broadening the application range of HTL technology. Given the current waste handling infrastructure worldwide, combined HTL (co-HTL) is of great interest due to its versatility and capacity to cope with the challenges posed. The combination of complex materials, such as commercial synthetic polymers and biomass, can result in unexpected interactions that must be explored and understood prior to full scale applications. So far, previous research within this topic is limited and few polymers have been investigated.

The first study within this scope reported co-HTL of mixtures of lignite, wheat straw and polyethylene terephthalate (PET), it showed that at certain ratios, the total conversion to oils of mixed materials was higher than the single-component liquefaction calculated prediction.[12] Recently, the lignocellulosic biomass, *Prosopis Juliflora*, was catalytically (activated bentonite) co-liquefied with non-identified polyolefin waste and reported to have the best oil yield (61.2%) using 3:1 (biomass:polyolefin) mixing ratio at 420 ºC (supercritical conditions) and 60 minutes reaction time. The same study reports positive synergies for bio-oil production in sub and supercritical water (340-440 ºC).[13] Using the same catalyst and biomass, co-HTL of paint sludge (described as containing synthetic polymer waste) was also reported to increase the oil-yield significantly using 2:1 ratio (biomass:paint sludge) for all investigated temperatures (340-440 ºC).[14]

Subcritical (350 ºC) co-HTL of pistachio hulls and individually PE, PET, PP, and nylon-6 (feed containing 10-20 wt. % plastics) was reported to achieve synergy effects in oil yield as well. The biocrude higher heating value (HHV) increased in mixtures of 10 and 20% PE, while it was maintained for the other polymers tested. However, the only polymer to yield a higher chemical energy recovery when compared to the pure biomass (60%) was PET at 10-20% mix ratio (70%).[15]

The present study aims to expand the state of knowledge and critically evaluate the technical feasibility and efficiency of co-HTL of synthetic polymers and the model lignocellulosic *Miscanthus Giganteus. A* systematical evaluation of 12 different synthetic polymers is presented, including previously unreported ABS, Epoxy, PA66, PC, PS and PUR, and different mixture ratios than current literature are reported for the all others. Unified process conditions, resembling those of continuous processing pilot reactors[16,17] were chosen to allow comparison amongst samples and identification of synergetic and antagonistic effects of different polymers with miscanthus. We also highlight how other important streams (aqueous phase, gas and solids) of HTL processing behave under such conditions. The study clarifies which reaction pathways and product effects we can expect to find upon co-HTL, highlighting opportunities and challenges of this approach.

## 2. MATERIALS

*Miscanthus Giganteus* was harvested at Aarhus University's facilities (Foulum, Denmark), extruded using a twin screw extruder (Xinda, 65 mm twin screw extruder with 2000 mm barrel length) and dried for 24h at 105 ºC in a forced circulation oven. A total of 12 different commercial polymers were investigated: poly-acrylonitrile-butadiene-styrene (ABS), Bisphenol-A based Epoxy resin, high density polyethylene (HDPE), low density polyethylene (LDPE), non-colored plastic cable ties of polyamide 6 (PA6), Sigma-Aldrich polyamide 6/6 (PA66), polyethylene terephthalate plastic bottles (PET), polycarbonate (PC), polypropylene cups (PP), polystyrene cups (PS) and polyurethane foam (PUR). Table S1 show their molecular structures. Most polymers were milled individually using a Polymix® PX-MFC 90D knife mill equipped with a 2 mm sieve prior to HTL, while Sigma-Aldrich Poly(vinyl chloride) (PVC) was used as acquired (powder).

## 3. METHODS

**HTL procedure.** Custom made Swagelok© bomb-type reactors were assembled as previously described[18] with a total volume of 20 mL. Experiments were conducted using a mixture of miscanthus and synthetic polymers of ratio 1:1 (mass basis) for each individual polymer. Reactors were loaded with 0.50 g of miscanthus and synthetic polymer and 8.50 g of water, sealed and submerged in a pre-heated fluidized sand-bath for 20 minutes at 350 °C. After reaction time, reactors were quenched in water, the mass of gas was determined gravimetrically by venting the room-temperature reactor, and the product work-up procedure followed. The aqueous phase (AP) was decanted into a centrifuge tube (spun for 5 minutes at 4000 rpm), while the remaining solids and viscous oil in the reactor were washed-up with 30 mL of methanol and vacuum filtered. Solids recovered from the bottom of the AP centrifuge tube were washed with methanol and filtered together with the reactor's content. Solids were collected and dried overnight in a 105 °C forced-circulation oven. An aliquot of 1 mL out of the 30 mL of methanol used for washing the reactor was used for gas chromatography coupled with mass spectrometry (GC-MS) analysis and the remainder was evaporated under light nitrogen flow to determine the biocrude weight after a minimum of 24 h. Yields are expressed in dry ash free basis (Equation 1).

Equation 1 $$Y_{experiment} = \frac{W_{phase}}{W_{feed}}$$

Synergy effects (SE) were determined according to Equation 2.

Equation 2 $$Synergy\ effect = \frac{Y_{experiment}}{X_{polymer}Y_{polymer} + X_{miscanthus}Y_{miscanthus}}$$

Where $Y_{experiment}$ represents the mass yield in $g/g_{feed}$ of a certain phase (oil, gas or solids – aqueous by difference), X is the fraction of material in the feedstock (0.5) and $Y$ is the yield from experiments with pure polymers[11] from previous data published by the authors or pure miscanthus.

**Elemental analysis.** Using an Elementar vario Macro Cube elemental analyzer (Langenselbold, Germany), Carbon, Hidrogen, Nitrogen and Sulphur (Oxygen by 100% difference) contents were determined for all raw materials, solid residues and oil products in duplicate, average values are reported. High heating value (HHV) of the fractions analyzed was estimated by Channiwala-Parikh correlation (Equation 3).[19] The energy yield in the oil phase of each experiment was calculated according to Equation 4, while the carbon yield is derived from Equation 5.

Equation 3 $$HHV\left[\frac{MJ}{kg}\right] = 0.3491\,C + 1.1783\,H + 0.1005\,S - 0.1034\,O - 0.0151\,N - 0.0211\,A$$

Equation 4 $$Energy\ yield\ \% = \frac{HHV_{oil}\left[\frac{MJ}{kg_{oil}}\right].Yield_{oil}\left[\frac{kg_{oil}}{kg_{feed}}\right]}{HHV_{feed}\left[\frac{MJ}{kg_{feed}}\right]} \times 100$$

Equation 5 $$Carbon\ yield\ \% = \frac{C_{oil}\left[\frac{kg_C}{kg_{oil}}\right].Yield_{oil}\left[\frac{kg_{oil}}{kg_{feed}}\right]}{C_{feed}\left[\frac{kg_C}{kg_{feed}}\right]} \times 100$$

**GC-MS.** An Agilent 7890B GC coupled to a quadrupole mass filter MS Agilent, 5977A was used for all analyses. For oil analysis, 1.0 µL out of the 1 mL methanol aliquot retrieved from sample work-up was diluted appropriately (from 1 to 8x) and internal standard was added (4-bromotoluene) prior to direct injection (inlet temperature of 280 °C, split ratio 20:1, helium flow 1 $mL.min^{-1}$). The GC column used was a VF-5ms column with dimensions 64.9 m x 0.25 mm x 0.25 µm, which experienced an oven temperature program initiating by 60 ºC hold for 2 minutes, a ramp to 200 ºC (5 $ºC.min^{-1}$), another ramp to 320 ºC (20 $ºC.min^{-1}$) and a final hold of 5 minutes at constant temperature. Compounds were identified with authentic standards, NIST17 mass spectra library or based on literature references.

Equation 6 was used to estimate the change in oil composition of polymer and miscanthus blends in comparison to HTL of pure feedstock materials (pure miscanthus or pure polymer). $A_{i,exp}$ is the area of compound *i* identified in the GC-MS of a certain experiment (pure miscanthus, pure polymer or blend) and it is used to discuss the change in oil composition, not the total conversion of the feedstock to a certain compound, as this would require complete quantification of each compound analyzed.

Equation 6 $$\frac{C_{i,exp}}{C_{i,max}} = \frac{A_{i,exp}}{A_{i,max}}$$

## 4. RESULTS AND DISCUSSION

### 4.1. MASS BALANCE

Figure 1 depicts the mass balance for all co-HTL experiments ordered by oil yield. Miscanthus' HTL is also included as reference for interpretation. Notably, oil yields vary from 10.1 to 43.2%, respectively co-HTL of LDPE and PUR. Most co-HTL experiments show significant decrease of the total oil yield, which is expected based on results previously published[11], given most of them do not decompose into oil directly. Namely, LDPE, HDPE, PP, PS and ABS yield around 50% solid materials, essentially composed of unreacted or charred polymeric material and only about 10% oil. This indicates lack of depolymerization for the co-HTL conditions tested. As for PA66, Epoxy, PC and PUR, the oil yield is improved and generally gas yields reduced, when compared to pure miscanthus HTL. It is noticeable that all polymers containing significant amount of nitrogen in the form of amines are located in right side of Figure 1, while ABS, which contains nitrogen in form of nitrile, appears towards the left side, together with polyolefins and PS.

PC is the only material that does not contain nitrogen and is located on the right side of Figure 1. HTL of PC alone has shown the highest oil yield among all individual polymers tested at the same condition[11], which can be the reason for its position among the highest oil yields of co-HTL as well. The mass balance shown in Figure 1 does not provide information if the co-liquefaction results in positive or negative synergies compared to what could be expected from their individual processing. Hence, mass yield synergy effects are depicted in Figure 2 to improve understanding of all co-HTL experiments

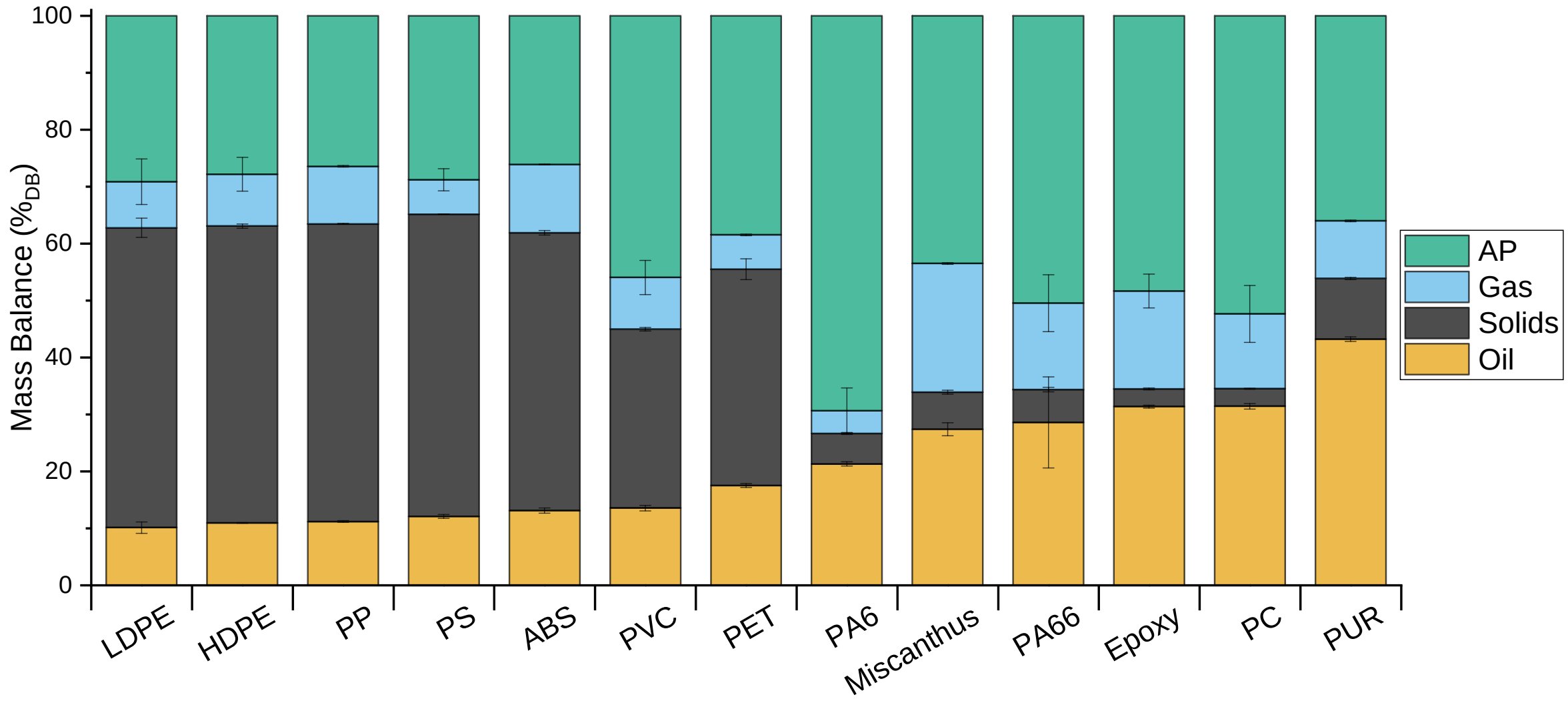

Figure 1 - Mass balance of co-HTL of synthetic polymers and miscanthus (1:1 mixing ratio) [AP by difference]

## 4.2. SYNERGY EFFECTS

The co-liquefaction synergy effects, based on previously published data[11] and the data of this study, shown in Figure 2 A-D depict several deviations from predicted results, demonstrating that products of miscanthus and synthetic polymers interact significantly during HTL processing. Figure 2A also depicts the nitrogen content of the raw synthetic polymer, showing a clear correlation between nitrogen presence and oil yield synergy effect. Polymers containing a larger amount of nitrogen tend to have positive synergy effects (>1, e.g. PUR at 1.54 and PA66 at 1.62), as nitrogen-containing organic molecules and ammonia under hydrothermal conditions act as catalysts of oil formation.[20] Surprisingly, PC depicted the lowest synergy effect for oil among all polymers tested, with a negative synergy of 0.63. This indicates that the products of PC HTL (mainly composed of bisphenol-A, p-isopropenyl phenol and phenol[11]) are reactive towards miscanthus-derived biocrude, yielding subsequent water-soluble compounds, evident from the large synergy of 1.65 for the aqueous yield (Figure 2D). PA6, PUR and PA66 present the highest oil synergy effect, while also having the highest nitrogen content among all polymers tested. PUR, however, presents a higher synergy effect than PA6, indicating that nitrogen is not the only factor interfering in the oil formation for these materials. It is thus clear, that even though nitrogen-containing polymers show a better chance to favor oil formation in the presence of miscanthus, the chemical characteristics of these materials play an important role in the result.

Figure 2B shows the synergy effects for solids. Antagonistic effects (SE < 1) for solids are observed for PA6, PA66 and Epoxy, all feedstocks depicting nitrogen concentrations higher than 4%, following the expectation of increased reactivity by ammonia formation. PC also has a similar antagonistic effect for solid formation as Epoxy (its thermosetting counter-part), even though it does not contain significant amounts of nitrogen, which again shows that the nitrogen content is not the only parameter affecting the reactions. PUR showed the highest solid synergy effect among all polymers tested, with a total solid yield for the co-liquefaction experiment of 10.7 % (Figure 1), despite containing significant amounts of nitrogen. Given these observations, it can be concluded that the chemical reactivity of nitrogen (determined by its conformation in the synthetic polymer) is more important than the presence of nitrogen for the synergistic effects observed. PVC does not exhibit the highest synergy effect on solid yield, which could be expected due to acidification of the media by HCl formation and consequent carbonization.[11,21] Despite PVC showing a positive solid synergy effect, PS and PUR are more prone to increase the solids content upon interaction.

Figure 2C shows the gas phase synergy effects, which are generally negative (<1) for most polymers, being positive only for PA66 and Epoxy. PA6 and PA66 differ in gas synergy effect by 0.96, the former depicting an antagonistic effect and the latter synergistic, which may be related to differences in nitrogen reactivity depending on the type of molecular structure.[20] PA6 HTL yields mainly secondary amides, while PA66 results in amines and amides[11], it is possible that the higher reactivity of the latter results in increase of gas yield via interactions with miscanthus-derived components. PA66 and Epoxy are the only co-HTL experiments with positive synergy effects for gas yields (PA66 and Epoxy), both containing amines as single-polymer HTL products.[11]

PET has a negative synergy effect for gas formation (Figure 2C), which indicates that the carboxyl groups present in biomass prone for gas formation tend to recombine into other products in the presence of PET. This can also be verified in Figure 4A, where PET oil is depicted having a higher O/C in comparison to pure miscanthus oil, which indicates an oil containing more carboxyl groups.

Figure 2D depicts the synergy effects for AP yield, which for most polymers is neutral. PUR shows a negative and PET, PA6 and PC a positive synergy effect. The negative synergy effect for PUR AP and gas products indicates migration of the products to oil and solids phases, both have pronounced positive synergy effects. As PET only presents deviation from neutral synergy in the gas phase, being negative, and in the AP, being positive, the data indicates that PET products (mainly composed of terephthalic acid and ethylene glycol[11]) tend to release less carbon in the form of $CO_2$ when miscanthus products are present in the media, giving preference for AP yields.

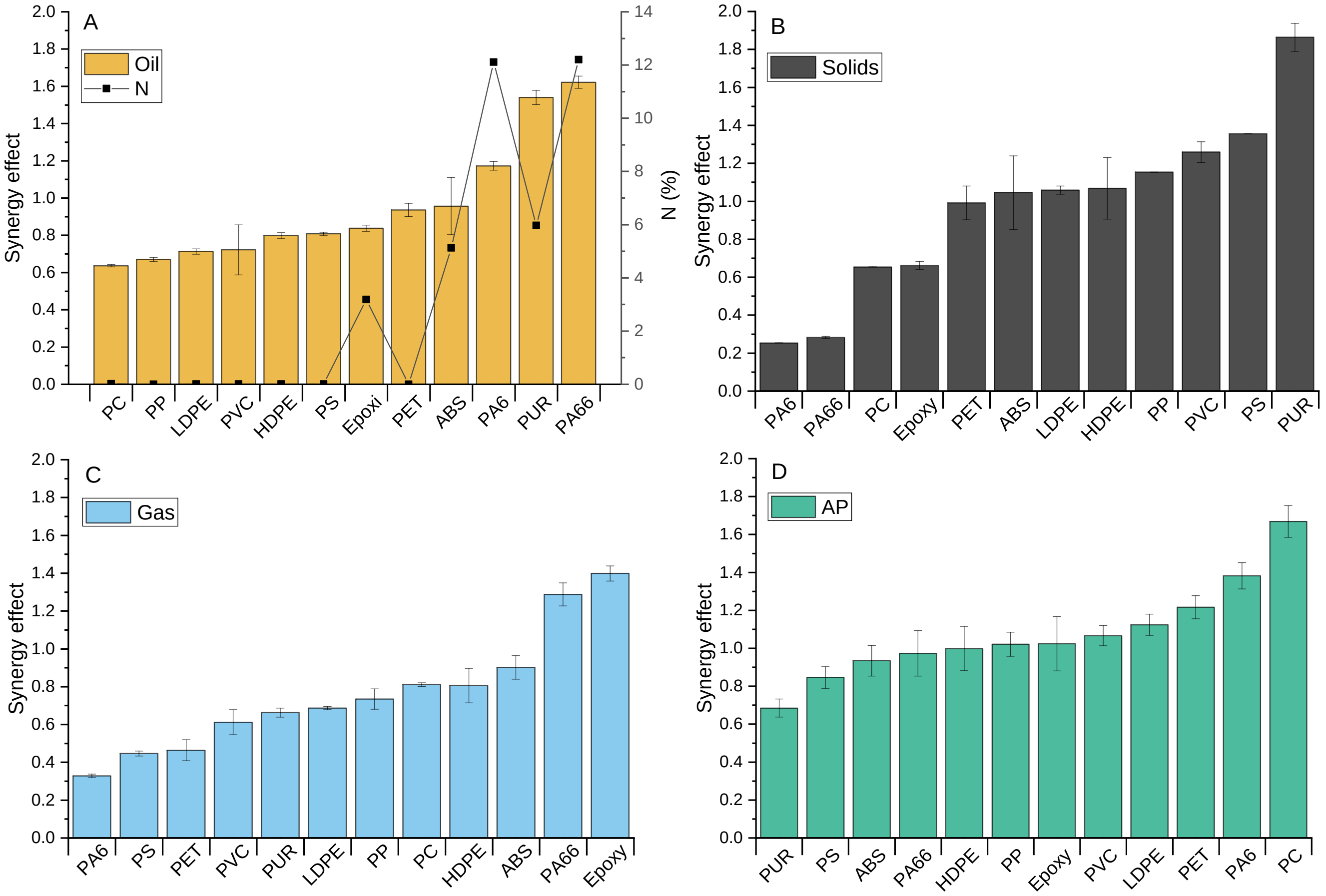

Figure 2 – Weight yield based synergy effects of synthetic polymers and miscanthus co-HTL (1:1 mixing ratio)

## 4.3. OIL YIELDS AND ATOMIC COMPOSITION

Figure 3 shows the oil energy, oil carbon yield and HHV for all experiments, plotted in an ascending order. The HHV of PVC was not calculated, as oxygen content was determined by difference and in PVC samples the difference is composed of both chlorine and oxygen. The figure shows that both oil energy and carbon yields have a very similar trend, ranging from 9.2% (LDPE co-HTL) to 55.3% (PUR co-HTL) and from 7.8% (PVC co-HTL) to 52.2% (PUR co-HTL), respectively. The HHV for miscanthus biocrude was calculated to be 29.3 MJ/kg. Among the co-HTL oils, LDPE, HDPE, PP, PS, PA6, PA66 and PUR are within 2.5% of the miscanthus biocrude HHV, while ABS, Epoxy and PC present values increased by of 6.9, 8.8 and 12.3% respectively, PET HHV is the only co-HTL oil to present a lower HHV than miscanthus biocrude, being 12.0% lower. PUR is the only synthetic polymer in co-HTL that increases oil energy and carbon yields in comparison to pure miscanthus HTL. This effect is not only observed because this polymer has one of the highest synergy effects for oil formation (Figure 2A), but also due to an increase in H/C ratio (see Figure 4A). It is also possible to see that PC and Epoxy are positioned side by side, with very similar energy and carbon yields to miscanthus, but higher HHV for both. The similar molecular structures of these two materials (see Table S1) result in similar HHV and energy yield in co-HTL, indicating that the overall interactions among lignocellulosic- and PC-like-derived biocrude are equivalent, even though synergy effects for the different product streams differ greatly (Figure 2 A-D).

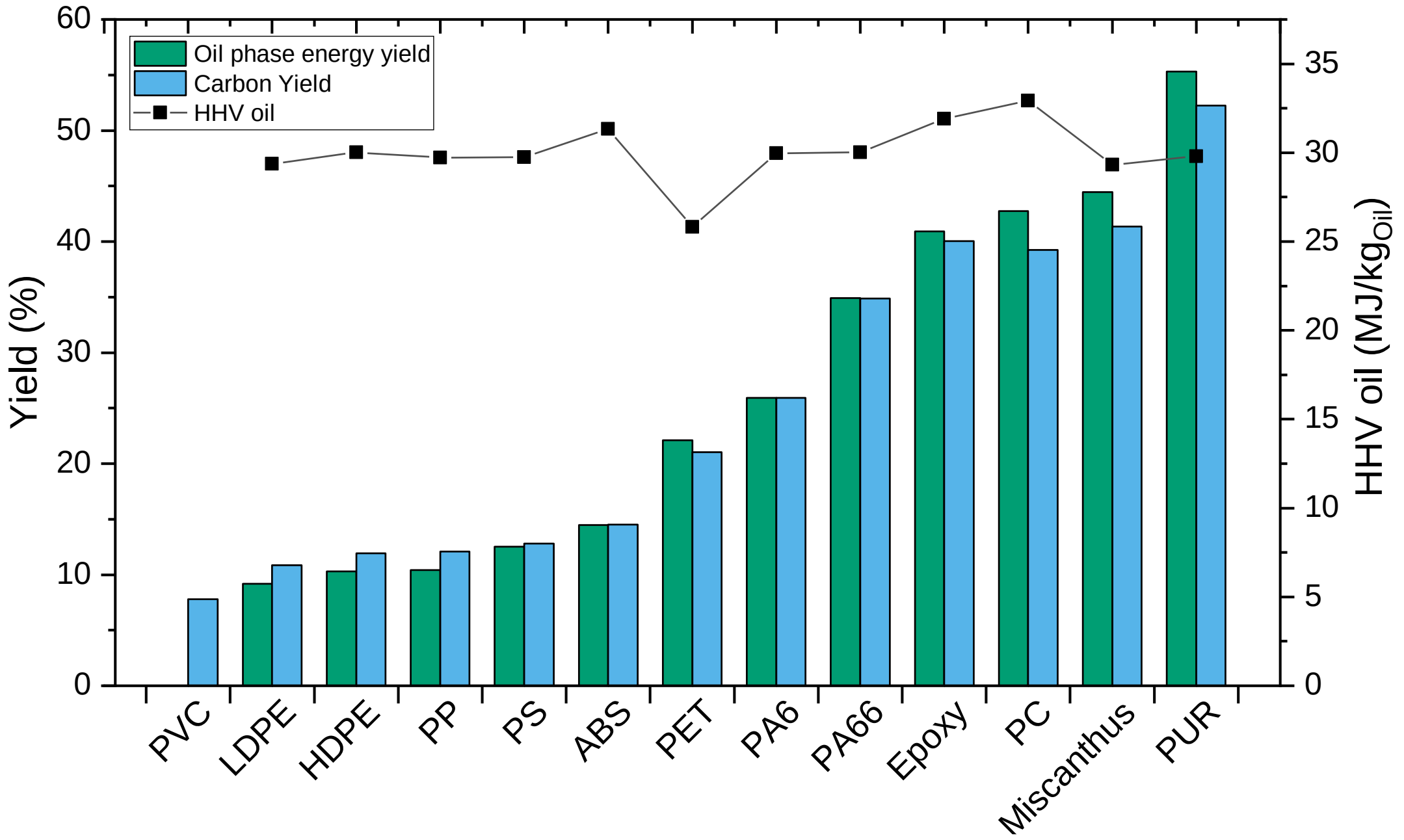

Figure 3 – Oil phase energy yield and HHV of synthetic polymers and miscanthus co-HTL (1:1 mixing ratio)

All polyolefins, PS and ABS depict very low energy and carbon yield due to the increase of both $HHV_{feed}$ and $C_{feed}$ without reciprocal increase in co-HTL oil yield (see Equation 4 and Equation 5 as reference). ABS has a higher HHV than the other products due to lower O/C (Figure 4A), which can be linked to the appearance of aromatic-like compounds in the oil composition (presented in Section 5).

The already discussed increase of O/C ratio in PET co-HTL oil is the reason for its HHV being lower than pure miscanthus biocrude, which also contributes to lower energy yield in comparison to other polymers. In the literature, PET co-HTL with pistachio hulls (10-20$\%_{w/w}$ PET in feed) has been reported to increase both oil phase energy yield and HHV[15], however our observations with miscanthus co-HTL show the opposite, indicating that feed composition can be optimized.

Co-HTL of PA6 gives a lower energy yield than PA66, even though both biocrudes have equal HHV. The difference in oil synergy effects of PA6 and PA66 components with miscanthus biocrude indicates that primary amines from PA66[11] tend to contribute to oil phase compounds, increasing energy yield, while keeping H/C, O/C and N/C ratios very similar (see Figure 4), resulting in a similar HHV for both.

Figure 4 A shows the van Krevelen diagrams (H/C versus O/C ) for all oil and feedstock samples while Figure 4B shows the H/C versus N/C data for experiments with significant amount of nitrogen in the feed composition. The polymer feed ratios plotted in the graph are based on a 1:1 mixture of polymer with miscanthus. When comparing the co-HTL feed mixture with pure miscanthus, it is possible to see that O/C and H/C ratios are very similar for all polyolefins and PS, indicating little contribution of the synthetic materials to biocrude formation. As for H/C ratios, PUR, PA6 and PA66 (labeled as PA in Figure 4 as points are too close to one another) have higher values than pure miscanthus, while PC, ABS and Epoxy also have higher values, though less pronounced.

With miscanthus biocrude as reference, it is possible to observe that PET is the only co-HTL oil product that gives both an increase in O/C ratio and slight decrease of H/C, which indicates that PET products of HTL tend to carry a large portion of oxygen into the oil phase in co-HTL with miscanthus. This conclusion is in agreement with the HHV observations previously discussed. Polyolefins and PS seem not to interfere in H/C and O/C ratios, which is not surprising as they do not contribute to oil formation. ABS, PC and Epoxy decrease O/C and increase H/C slightly. Both PA and PUR maintain the same O/C while increasing H/C compared to pure miscanthus biocrude. These observations, in combination with the results of oil phase synergy effects (Figure 2), show that not only oil yields are increased, but their quality regarding H/C is also improved.

Figure 4 B shows only experiments involving nitrogen-containing synthetic polymers. It is possible to observe a decrease of N/C ratio for all oil products when compared to the original co-HTL feed. Generally, the higher the feed N/C ratio, also the higher the oil product N/C. This does not occur when comparing Epoxy to ABS. This last observation may be attributed to the type of nitrogen group present in both raw materials, the former being a secondary or tertiary amine, the latter being a nitrile group. It seems, thus, that nitrogen in the form of nitrile groups is not as prone as amines to migrate to biocrude in co-HTL with lignocellulosics. That observation follows the lack of reactivity of ABS when processed via HTL by itself.[11]

The N/C ratios of PUR and PA oils (Figure 4 B) are closer to those of their respective feeds than to pure miscanthus biocrude. This indicates high efficiency of transfer of nitrogen from feedstock to biocrude during co-HTL. In Figure 2A, PUR and PA oil phase positive synergy effects are prominent, thus it seems that nitrogen in the form of amines contributes to oil yield increase, possibly participating in the formation of more hydrophobic products partitioning to the oil phase.

This observation is corroborated with results indicating increase of oil yield for the co-HTL of nitrogen-containing biomass and low nitrogen lignocellulosics.[7–9,22,23] Previous results show that co-HTL of different lignocellulosics – including miscanthus – with sewage sludge, manure or algae has the effect of improving biocrude yields[22,24]. Similar effect has been observed with co-HTL of algae and wood.[25] Lignin and *Spirulina platensis* co-HTL using a feed ratio of 2:1 also results in positive synergy effects for oil yield[8], as well as *Enteromorpha clathrata* and rice husk.[7] In another study using model compound mixtures, all protein containing co-HTL with cellulose or lignin derived materials showed higher biocrude yields than expected.[23] In all literature examples, biomasses containing significant amount of nitrogen in the form of amines result in increase of biocrude yield. The present study reports for the first time similar findings for synthetic polymers. However, two of the materials studied here are examples of nitrogen-containing polymers that do not result in significant positive synergy effects for oil formation: ABS and Epoxy. Both also do not contain nitrogen in the form of primary or secondary amines, but rather as nitrile groups or tertiary amines. Thus, the chemical conformation of nitrogen is important for synergy effects to be positive in co-HTL processing of lignocellulosics and nitrogen-containing materials, with higher synergy effects for oil formation occurring in presence of nitrogen in the form of secondary or primary amines.

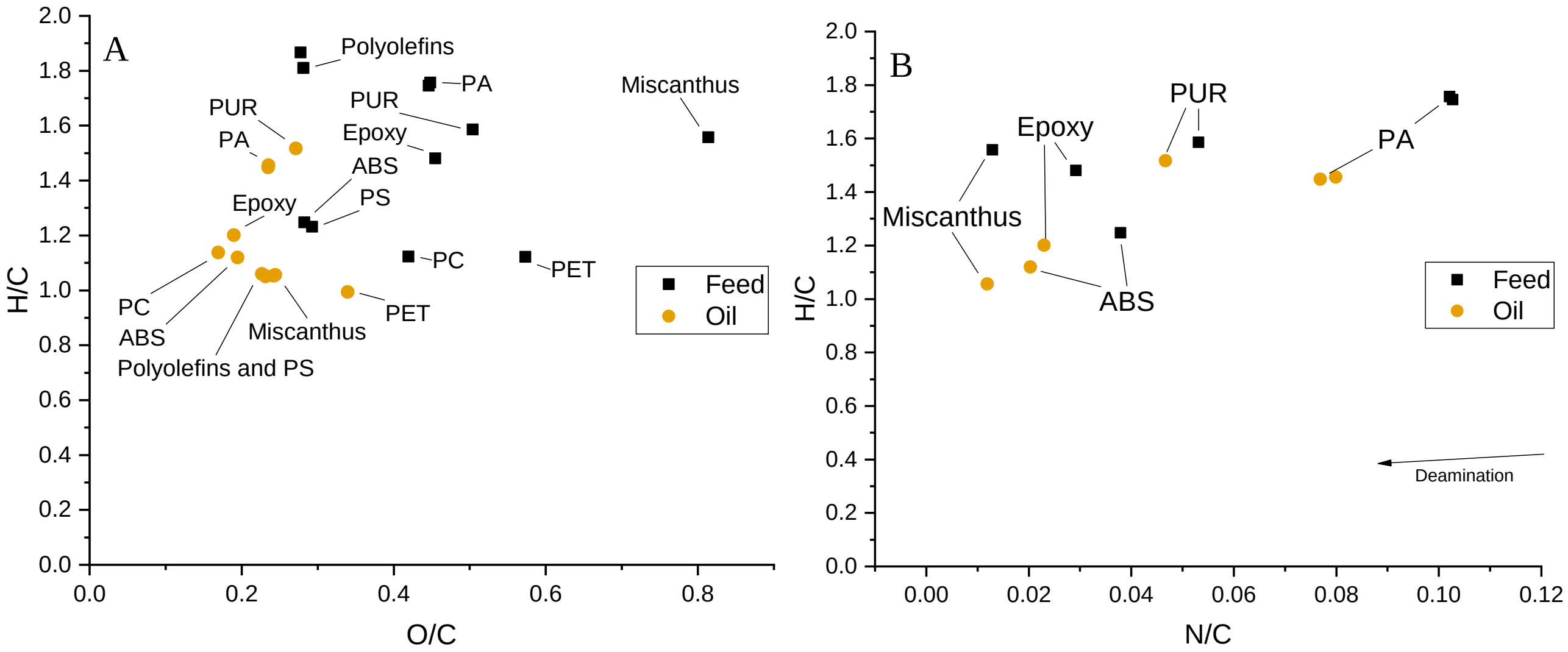

Figure 4 – H/C-O/C (A) and H/C-N/C (B) van Krevelen diagrams of synthetic polymers and miscanthus co-HTL (1:1 mixing ratio feed and oil product) (PA6 and PA66 label as PA)

## 4.4. COMPOSITION OF CO-HTL OILS

The changes in relative concentration of specific compounds measured by GC-MS can help understand how the composition of the volatile fraction of biocrude changes according to the feedstock used. It is worth noting that the following section does not refer to the amount of a certain compound that can be recovered from a certain blend, rather it explores how concentrations of compounds change in the presence of synthetic and lignocellulosic materials. For a better understanding, the next sections are grouped according to similarity of the synthetic polymers used in co-HTL experiments.

### *4.4.1. LDPE, HDPE, PP, PS and ABS*

Figure S1 shows that selected methoxylated compounds are found at higher concentrations when miscanthus is co-liquefied with polyolefins. At the same time, Figure S2 shows that o-methoxy phenols relative concentrations are not changed, while alkyl o- and p- phenolics are increased when HTL takes place in presence of polyolefins. This indicates that the presence of polyolefins favors the formation of alkyl phenols, though it does not interfere significantly in the formation of methoxy-phenols. This observation points out that alkyl-groups are more prone to be added to phenolic structures derived from lignocellulosic biocrude when polyolefins are present in HTL media. Table S2 depicts that, even though phenolics are still the major components found in co-HTL of HDPE and LDPE, ethers and acids become more prominent for those in comparison to pure miscanthus biocrude. The same data for PP shows a high prominence of tetrahydrofuran (21.46%) together with aromatic compounds (e.g. 1,2,3-trimethoxy-5-methyl-Benzene, 1.91%; 7-Methoxy-1-naphthol, 1.60%). PP and pistachio hulls co-HTL has been described to change the bio-oil composition significantly[26], however not specifically identifying the compounds involved in this change. The observations above indicate that the compounds formed in co-HTL of PP are more likely to change the composition of the biocrude in comparison to HDPE and LDPE.

PS and ABS also do not significantly affect biocrude yields, though presence of phenolics in the product is increased (except for 2-methoxy-phenol), together with several different single-substituted aromatics (see Figure S3 and S4). For all five polymers (LDPE, HDPE, PP, PS and ABS) meta-alkyl substituted compounds were not present. This is directly linked to the reactivity of phenolic compounds derived from lignocellulosic biomass, which gives preference for ortho and para alkyl substitutions.[27] The GC-MS area-based composition of biocrude from co-HTL of ABS shows a high presence of styrene, ethyl benzene and benzene-butanenitrile, structures very similar to the monomers of ABS (Table S1 and Table S2). For PS, Table S2 shows a distinct presence of 3-(2-cyclopentenyl)-2-methyl-1,1-diphenyl-1-propene and similar isomers, also to a lesser extent of styrene, ethylbenzene and other polyaromatics. The presence of these compounds with similar structure in comparison to the monomers of the synthetic resins used in co-HTL is interesting for the valorization of biocrude as a source not only of fuels, but possibly as a source of platform chemicals.

The observations above indicate that synthetic resins without backbone heteroatoms affect the quality of miscanthus biocrude, despite not depolymerizing into oil products under these conditions, corroborating the van Krevelen diagram findings discussed previously (Figure 4). This change in quality occurs mainly due to the addition of alkyl groups in biocrude volatile molecules. The occurrence of such alkyl groups in the backbone of the polyolefins co-processed creates a potential opportunity for biocrude quality improvement with this co-HTL strategy despite the lack of oil yield increase.

#### *4.4.2. PVC*

Comparison of PVC co-HTL derived oil with miscanthus biocrude (Table S2) shows a remarkably higher presence of carboxylic acids, aromatics and polyaromatics. This is expected based on acid-catalyzed reactions in lignocellulosic biomass HTL. Chlorine released from the polymer side group hydrolyses into HCl, acidifying the media and promoting dehydration reactions of the phenolic compounds, yielding aromatics. For PVC, hydrothermal processing instead yields a Cl poor solid phase, concentrating much of the chlorine into the aqueous phase.[21] Low temperature hydrothermal co-processing of PVC and lignocellulosic biomass has been proven to be an effective method for dechlorination.[21] Higher subcritical temperatures, such as that tested in this study (350 °C), do not provide advantages regarding the generation of a valuable oil phase.

*4.4.3. PC and Epoxy*

The co-HTL oil produced from PC and Epoxy have very similar trends, as shown in Figure 5. The concentration comparison shows very clearly that phenolic compounds are more prominent in co-HTL oil than in miscanthus or synthetic polymers alone. Phenolic compounds have partial solubility in water and are formed from HTL of both pure PC and Epoxy, which indicates that these are attracted to biocrude (a non-polar phase) when present in the reaction media, increasing the oil quality. The observation can also indicate that co-processing favors these compounds. The improvement in oil quality by co-processing is also depicted in Figure 4A by the decrease of O/C ratio in comparison to pure miscanthus HTL biocrude.

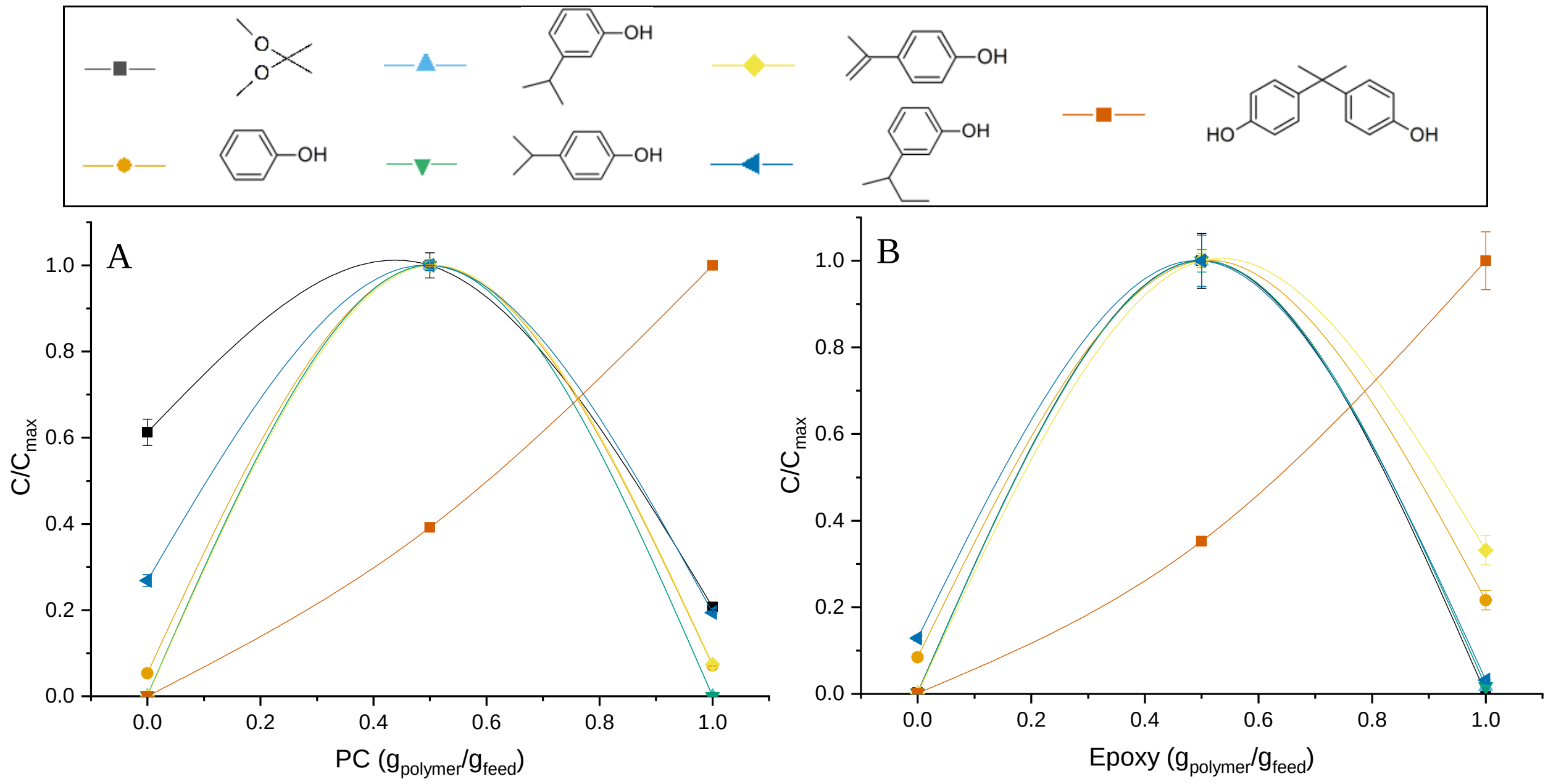

Figure 5 – Relative concentration change in PC (A) and Epoxy (B) co-HTL oil for selected compounds. (—■— 2,2-dimethoxy-Propane; —●— Phenol; —▲— 2-(1-methylethyl)-Phenol; —▼— 3-(1-methylethyl)-Phenol; —◄— 3-(1-methylpropyl)-Phenol; —■— 4,4'-(Propane-2,2-diyl)diphenol

4,4'-(propane-2,2-diyl)diphenol (bisphenol-A) follows the opposite curve, indicating a lower concentration than expected upon co-HTL. This fact partially explains the higher concentration of the other compounds, which are direct byproducts of bisphenol-A cracking or hydrolysis. Even though this is the case, it cannot fully explain the significant concentration increase of the other compounds, which may result from rearrangement of both miscanthus and PC (or Epoxy) components. We suggest two possible pathways to explain this observation. First, it is possible that bisphenol-A is relatively stable towards the media, thus, the increase in presence of other compounds is due to rearrangements involving products of both feedstock materials. Another possibility is that oligomers yielded from PC and Epoxy HTL are more easily converted to bisphenol-A in the presence of miscanthus HTL products, compensating the consumption of this compound in the formation of the byproducts and keeping concentrations proportional. The latter option follows chemical equilibrium principles, and recently published kinetics[28] suggests bisphenol-A is present in a high concentration around 20 min reaction time, thus the hypothesis explains the increase in byproducts of bisphenol-A hydrolysis.

### *4.4.4. PET*

PET co-HTL experiments yield an oil richer in phenols, ketones, and carboxylic acids (Figure S5). The terephthalic acid from the synthetic resin seems to contribute to this formation, however only a minor part of this component has reacted into different products, as its majority is present in the solid products and PET synergy effect for this phase is neutral (Figure 2). A significant conversion of terephthalic acid into oil components would be indicated by a negative synergy effect on solids, which is not the case.

The second major product of PET HTL is located primarily in the aqueous phase[11], ethylene glycol, however it is also partially found in the oil phase due to partitioning. It appears not to interact with the products of miscanthus liquefaction, as shown in Figure S5. The lack of reactivity of ethylene glycol towards lignocellulosic products of HTL has been reported before.[29] However, PET co-HTL oil has a very different volatile composition, as shown in Table S2. Here, we can observe a more pronounced presence of acids, methoxy- and carboxyl- containing compounds. This is in agreement with the findings of Figure 4. Despite lowering the energy content in the oil phase, such compounds may be of use for direct applications, being more valuable than its energy use.

### *4.4.5. PA6 and PA66*

Besides resulting in an increased oil yield, both PA6 and PA66 also change the oil composition significantly (see Figure S6 and Figure S7). In one hand, the presence of PA6 increases the concentration of cycloketones and mono-substituted phenols, while not changing double-substituted phenols. Given PA6 products of HTL are mostly composed of amines[11], it is unlikely that those would form phenolic or cyclic compounds without the presence of nitrogen in the composition if known mechanisms are followed.[27] Thus, it seems the media provides a better chance for the miscanthus-derived compounds to react into phenolics that are incorporated into the oil phase. On the other hand, PA66 seems to have the opposite effect for 2-methyl-2-cyclopenten-1-one and mono-substituted phenols. Even so, PA66 still promotes positive oil synergies and superior carbon yield if compared to PA6. This could be derived from larger compounds being formed from PA66- and miscanthus-derived compounds in comparison to PA6.

Table S2 provides the comparison between miscanthus biocrude and the co-HTL oil composition of PA6 and PA66. The PA-derived oils contain significantly more compounds with nitrogen, including the dimer 1,8-diazacyclotetradecane-2,7-dione, clearly derived from PA structures. Hexadecanoic acid, which is one of the monomers for PA66 production, is also present in these samples.

Given the composition of volatiles in the oil, it is possible to assert that the biocrude of co-HTL with PA is rich in monomer-like compounds which can possibly be recovered through co-HTL.

### *4.4.6. PUR*

PUR co-HTL with miscanthus had the most promising oil synergistic effect identified, being the only combination to increase both energy and carbon yields. The GC-MS discussion will try to suggest the reasons behind the superior yields. Figure 6 depicts the concentration changes for selected compounds measured by GC-MS. It is possible to observe in Figure 6A a similar concentration for ethyl benzene in all experiments, also only modest deviations for cyclopentanone and styrene. Cyclopentanone presents the highest concentration in miscanthus HTL, while styrene in pure PUR, though both have slightly lower concentrations than expected in the co-HTL oil. On the other hand, phenol seems to be more prevalent in the co-HTL than in the single-HTL oil.

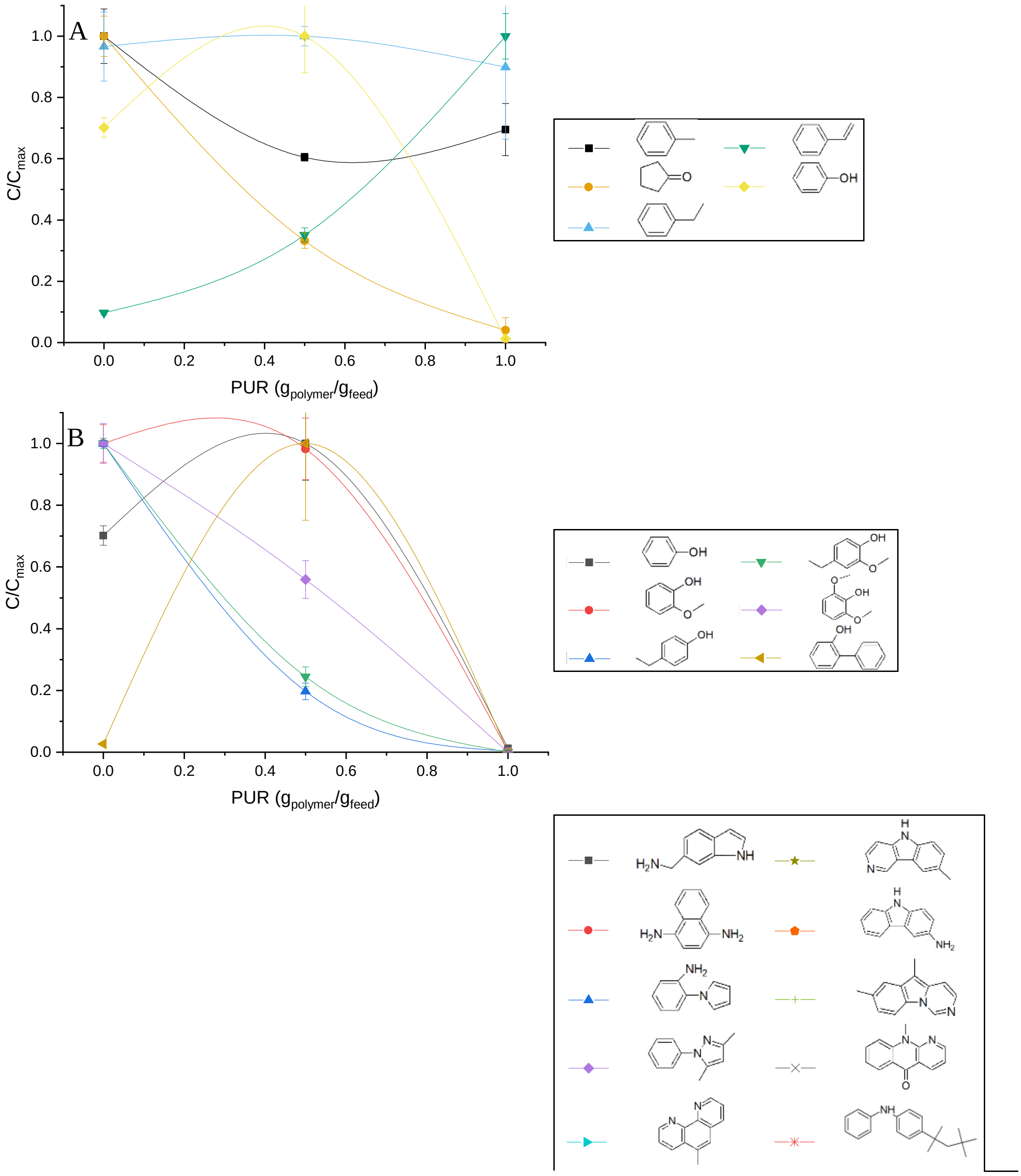

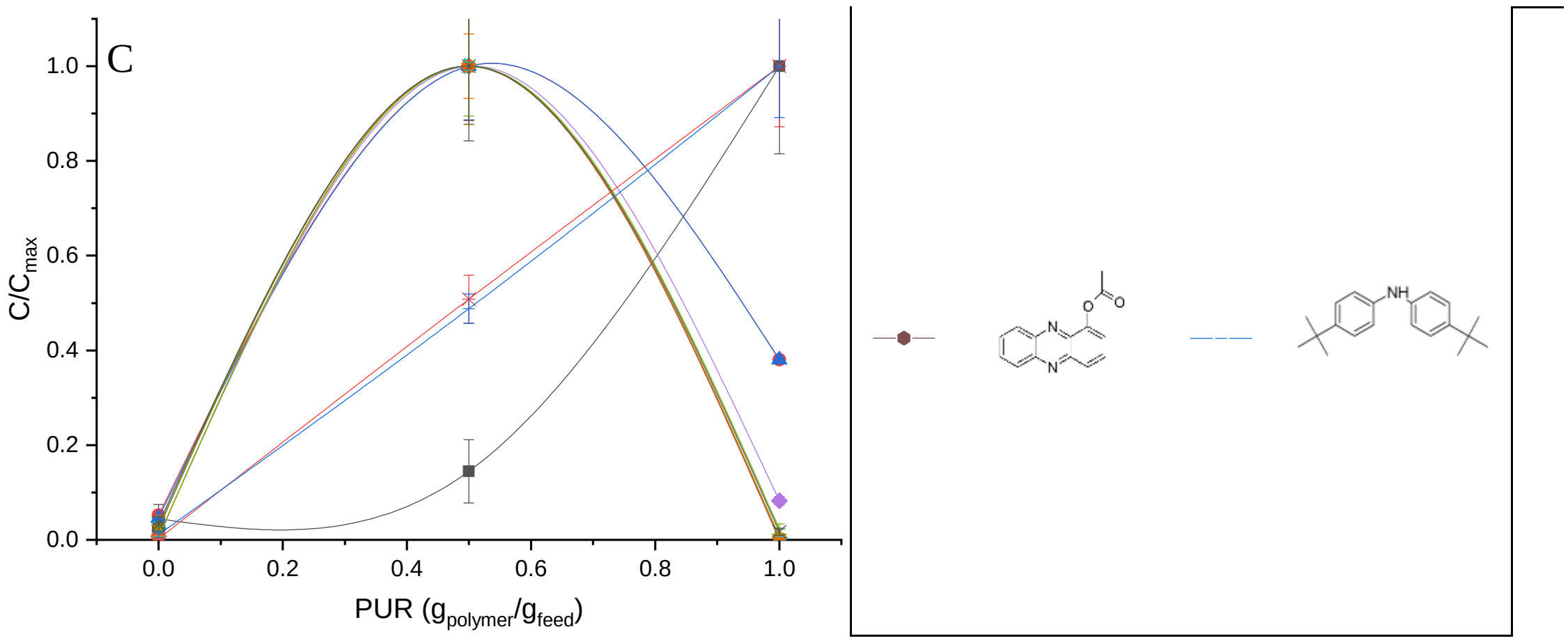

Figure 6 – Relative concentration change in PUR co-HTL oil for selected aromatic (A), phenolic (B) and nitrogenated (C) compounds. (A → —■— Toluene; —●— Cyclopentanone; —▲— Ethylbenzene; —▼— Styrene; —◆— Phenol); (B → —■— Phenol; —●—2-methoxy-phenol; —▲—4-ethyl-phenol; —▼—4-ethyl-2-methoxy-phenol; —◆—2,6-dimethoxy-phenol; —◀— o-hydroxybiphenyl); (C → —■— 1H-indol-6-ylmethanamine; —●— 1,4-Naphthalenediamine; —▲— 1-(2-Aminophenyl)pyrrole; —◆— 3,5-dimethyl-1-phenyl-1H-Pyrazole; —▶— 5-Methyl-1,10-phenanthroline; —●— acetate (ester) 1-Phenazinol; —★— 8-Methyl-5H-pyrido[4,3-b]indole; —⬟— 3-Aminocarbazole; —+— 5,7-Dimethylpyrimido-[3,4-a]-indole; —×— 10-methyl-Benzo[b]-1,8-naphthyridin-5(10H)-one; —✳— Tert-octyldiphenylamine; —— 4,4'-Di-tert-butyl-diphenylamine)

Figure 6B depicts changes in concentration of phenolics. Despite phenol presence being favored in co-HTL, followed by o-hydroxybiphenyl and, to a less extent, 2-methoxy-phenol, di-substituted phenols (4-ethyl-phenol and 4-ethyl-2-methoxy-phenol) have the opposite behavior. The tri-substituted 2,6-dimethoxy-phenol seems to be present proportionally in the co-HTL oil (around 0.5 in Figure 6B). Figure 6C depicts a group of compounds found with relatively small areas (the highest being 6,1% for 8-Methyl-5H-pyrido[4,3-b]indole – see Table S2), however in great abundance. These compounds are all heterocyclic aromatic containing nitrogen within the rings and all seem to be formed only in the presence of both PUR and miscanthus.

Table S2 also shows that the presence of 5 heterocyclic nitrogen-containing aromatics within the 20-most prominent identified compounds is the major difference between PUR co-HTL oil and miscanthus. The presence of such compounds indicate that the positive synergies and higher yields discussed before are connected to the reaction between PUR- and miscanthus-derived compounds, yielding this class of chemicals. Interestingly, Table S2 also shows that there are none of such compounds found within the 20 most prominent for all other nitrogen-containing tested materials (namely PA6, PA66, epoxy and ABS).

It is possible to observe by the nucleophilic reactivity of amines[30] and amides[31], that aromatic compounds with amine side groups are much more prone to react with phenolic compounds than organic molecules containing amides or nitriles. The two latter are the form of nitrogen present in the HTL products of PA6, PA66, epoxy and ABS, while the former is present in PUR.[11] This observation corroborates the discussion around Figure 2A, pointing towards the nature of the nitrogen-containing compounds being responsible for oil synergies. Figure 7 depicts a reaction scheme for PUR- with lignocellulosic-derived compounds, yielding products due to nucleophilic combinations, which is corroborated by other hypotheses previously described.[27] This is the first time nitrogen compounds from synthetic polymers are observed having behavior similar to protein in HTL conditions and interacting significantly with lignocellulosic-derived compounds. However, for the case of PUR, the oil formation effect is more pronounced due to the reactivity of the nitrogen species involved.

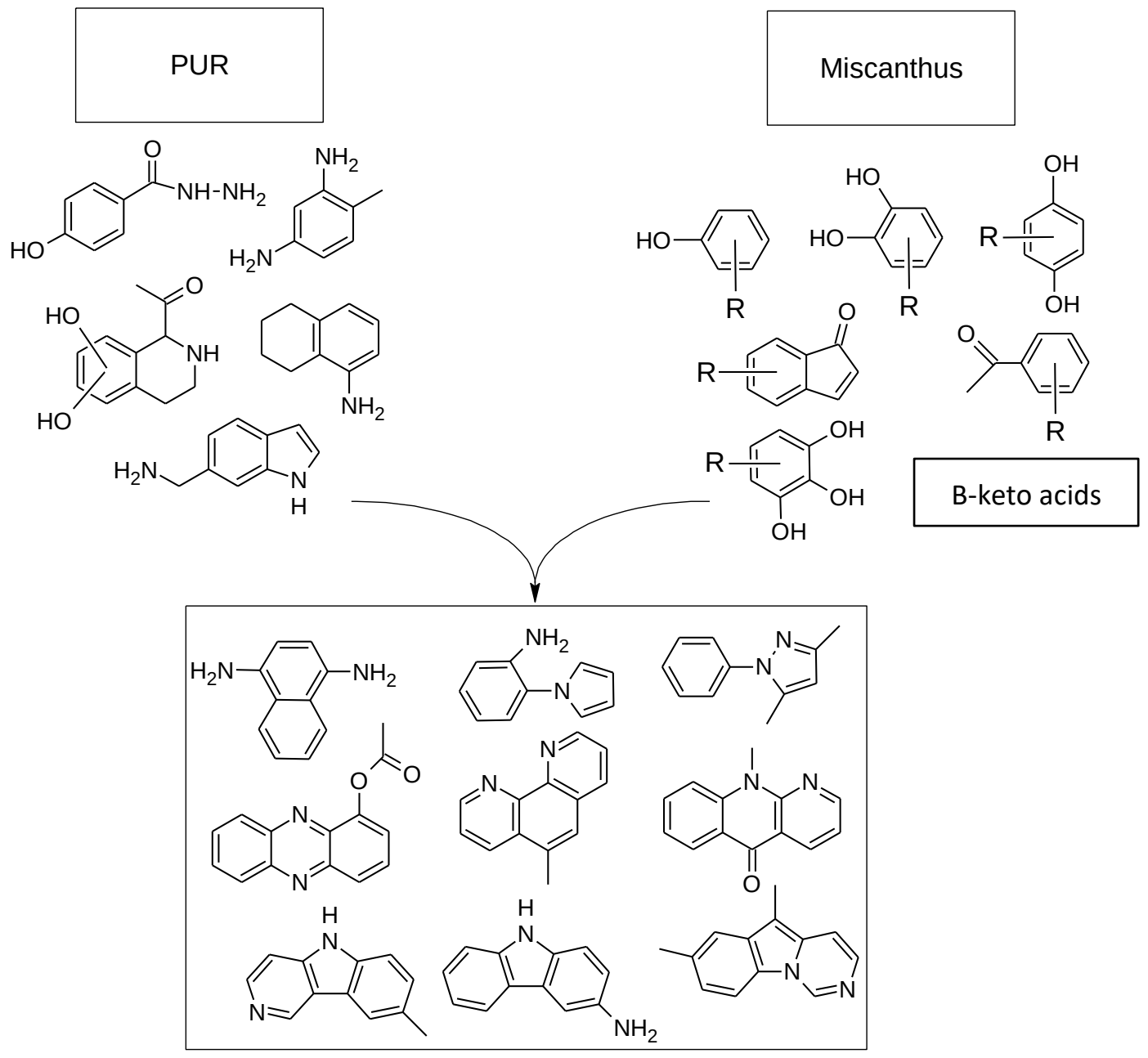

Figure 7 – Reaction scheme for PUR- and Miscanthus-derived HTL products.

## 5. CONCLUSION

Co-HTL of 12 different polymers with miscanthus resulted in different synergy effects and yields depending on the original synthetic polymers' chemical structures. A summary of the observations made through the present study is presented below:

- Positive oil yield synergy effects were observed from co-HTL of PA6, PA66 and PUR with miscanthus;
- Only PUR co-HTL with miscanthus increased carbon and energy oil yields when compared to single-HTL of miscanthus. PC and Epoxy maintained equivalent carbon and energy yields in the same comparison, while all other polymers decreased both parameters.
- HHVs of co-HTL oils from LDPE, HDPE, PP, PS, PA6, PA66 and PUR are within 2.5% of the miscanthus biocrude. ABS, Epoxy and PC, on the other hand, present increased values by 6.9, 8.8 and 12.3% respectively. PET HHV is 12.0% lower than miscanthus biocrude, due to the presence of more oxygenated molecules.
- ABS, despite having small synergistic effect, change significantly the oil composition, increasing biocrude quality, even though transfer of degradation products of these material to the oil phase is limited.

- PVC promotes severe carbonization reactions and the resulting acidic media leads to aromatics and polyaromatics formation in the oil phase.
- PC and epoxy co-HTL yields an oil rich in monomer-like structures of the original materials.
- PA6 and PA66 co-HTL oil is rich in nitrogen and also contains monomer-like structures, however to a lower extent.
- The type of nitrogen group in the synthetic polymer dictates the synergy effects. Amines connected to aromatics are more prone to recombine with lignocellulosic biocrude than alkyl amines, amides and nitriles. PUR, PA6, PA66 and ABS co-HTL synergies and oil compositions illustrate the findings.

## ACKNOWLEDGMENTS

This research was funded by the European Union's Horizon 2020 research and innovation program under grant agreement No. 764734 (HyFlexFuel – Hydrothermal Liquefaction: enhanced performance and feedstock flexibility for efficient biofuel production) and the Centre for Circular Bioeconomy (CBIO) of Aarhus University.